\documentclass{elsart}

\usepackage{graphics}
\usepackage{graphicx}
\usepackage{epsfig}

\usepackage{amssymb}

\usepackage{slashbox}

\usepackage{bm}

\begin{document}

\begin{frontmatter}

\title{Cyclic Codes from Two-Prime Generalized Cyclotomic Sequences of Order 6}

\thanks[label0]{This work was supported by the open fund of State Key Laboratory of Information Security, the Natural Science Fund of Shandong Province (No.ZR2010FM017), the National Natural Science Foundations of China(No.61170319)and the Fundamental Research Funds for the Central Universities(No.11CX04056A).}
\author[label1]{Tongjiang Yan}\ead{yantoji@163.com}
\author[label2]{Yanyan Liu}
\author[label3]{Yuhua Sun}
\address[label1]{College of Science, China University of Petroleum, Qingdao 266580, China.}
\address[label2]{College of Science, China University of Petroleum, Qingdao 266580, China.}
\address[label3]{College of Science, China University of Petroleum, Qingdao 266580, China.}

\begin{abstract}
Cyclic codes have wide applications in data storage systems and communication systems. Employing two-prime Whiteman generalized cyclotomic sequences of order 6, we construct several classes of cyclic codes over the finite field $\mathrm{GF}(q)$ and give their generator polynomials. And we also calculate the minimum distance of some cyclic codes and give lower bounds of the minimum distance for some other cyclic codes.
\end{abstract}


\end{frontmatter}

\newtheorem{lemma}{Lemma}
\newtheorem{definition}{Definition}
\newtheorem{theorem}{Theorem}
\newtheorem{remark}{Remark}
\newtheorem{corollary}{Corollary}
\newtheorem{proposition}{Proposition}
\newtheorem{proof}{Proof}
\newtheorem{conjecture}{Conjecture}
\newtheorem{example}{Example}

\section{Introduction}
Cyclic codes are an interesting type of linear codes and have applications in communication and storage systems due to their efficient encoding and decoding algorithms. They have been studied for decades and a lot of progress has been made, for example, some constructions and properties of them can be found in \cite{Ref1,Ref6,Ref7,Ref8,Ref10,Ref5,Ref9,Ref2,Ref3,Ref15,Ref4,Ref14,Ref12,Ref11,Ref13}.

Recently, several classes of cyclic codes using two-prime Whiteman generalized cyclotomic sequences and cyclotomic sequences of order 4 have been presented by C. Ding in \cite{Ref16} and \cite{Ref17} respectively, and lower bounds on the nonzero minimum hamming weight of some cyclic codes were developed at the same time. Y. Sun, T. Yan and H. Li presented several classes of cyclic codes using another class of two-prime Whiteman generalized cyclotomic sequences of order 4, and gave lower bounds on the nonzero minimum hamming weight of these cyclic codes \cite{Ref18}. The two-prime sequence is
employed to construct several classes of cyclic codes of order $d (d>0)$ over $\mathrm{GF}(q)$ and lower bounds on the minimum weight of these cyclic codes were
developed in \cite{Ref16}.
In this paper, we consider the cyclic codes over the finite field of order $q$ $\mathrm{GF}(q)$ from another class of two-prime Whiteman generalized cyclotomic sequences of order 6, and present the generator polynomials of these cyclic codes. Besides, we also calculate the minimum distance of some cyclic codes and give lower bounds of the minimum distance for some other cyclic codes.

Let $q$ be a power of any prime $p$, $n$ a positive integer satisfying $\gcd(n,q)=1$. A linear [$n,k,d$] code over $\mathrm{GF}(q)$ is a $k$-dimensinal subspace of $\mathrm{GF}(q)^n$ with minimum (Hamming) nonzero weight $d$. A linear [$n,k$] code $\bm{C}$ over the finite field $\mathrm{GF}(q)$ is called cyclic if $(c_0,c_1,\cdots,c_{n-1})\in \bm{C}$ implies $(c_{n-1},c_0,c_1,\cdots,c_{n-2})\in \bm{C}$ \cite{Ref6}. For any vector $(c_0,c_1,\cdots,c_{n-1})\in \mathrm{GF}(q)^n$, we identify it with the polynomial $c_0+c_1x+c_2x^2+\cdots+c_{n-1}x^{n-1}\in \mathrm{GF}(q)[x]/(x^n-1)$. Then any code $\bm{C}$ of length $n$ over $\mathrm{GF}(q)$ is equivalent to a subset of $\mathrm{GF}(q)[x]/(x^n-1)$. It is well known that $\mathrm{GF}(q)[x]/(x^n-1)$ is a principal ideal ring, i.e., any one of ideals of $\mathrm{GF}(q)[x]/(x^n-1)$ is principal. Let $\bm{C}=(g(x))$ be a cyclic code and $h(x)=\displaystyle\frac{x^n-1}{g(x)}$. Then we call $g(x)$ the generator polynomial and $h(x)$ the parity-check polynomial of $\bm{C}$ \cite{Ref6}.

Let $\bm{s^n}=(s_i)_{i=0}^{n-1}$ be a sequence of period $n$ over $\mathrm{GF}(q)$. We call
\begin{equation}\label{eq1}
S^n(x)=\sum_{i=0}^{n-1}s_ix^i\in \mathrm{GF}(q)[x]
\end{equation}
the generator polynomial of the sequence $\bm{s^n}$. It is well known that the minimal polynomial of $\bm{s^n}$ is given by
$$(x^n-1)/\mathrm{gcd}(x^n-1,S^n(x)).$$
Then the cyclic code $\bm{C_s}$ generated by the minimal polynomial of $\bm{s^n}$ is called the cyclic code defined by the sequence $\bm{s^n}$, which is a linear [$n,k$] code with $k=\deg(\gcd(x^n-1,S^n(x)))$. And the generator polynomial $g(x)$ of cyclic code $\bm{C_s}$ is equal to the minimal polynomial of the sequence $\bm{s^n}$, i.e.,
\begin{equation}\label{eq2}
g(x)=\frac{x^n-1}{\mathrm{gcd}(x^n-1,S^n(x))}.
\end{equation}
Correspondingly, the sequence $\bm{s^n}$ is called the defining sequence of the cyclic code $\bm{C_s}$ \cite{Ref6}.

Let $n_1$, $n_2$ be two distinct odd primes satisfying $\mathrm{gcd}(n_1-1,n_2-1)=6$. Let $e=\displaystyle\frac{(n_1-1)(n_2-1)}{6}$, $n=n_1n_2$ and $Z_n^*$ denotes the multiplicative group of integers modulo $n$. Suppose that $g$ is a common primitive root of $n_1$ and $n_2$ and $x$ an integer satisfying
$$x\equiv g(\mathrm{mod}\ n_1)\ \ \mathrm{and}\ \ x\equiv 1(\mathrm{mod}\ n_2).$$
Define
$$D_i=\{g^sx^i:s=0,1,\cdots,e-1\},\ i=0,1,\cdots,5.$$
Whiteman has proved that
$$Z_n^*=\bigcup_{i=0}^5D_i,\  D_i\cap D_j=\emptyset,$$
for $i\neq j$, where $\emptyset$ is the empty set \cite{Ref20,Ref21}. $D_i$ is called the two-prime Whiteman generalized class of order 6 for $i=0,1,\cdots,5$ \cite{yan2}.

Let $p,q,n,n_1,n_2,g,x$ be defined as before. Without special notation, $n_1,n_2$ always satisfy $\gcd(n_1-1,n_2-1)=6$. Note that $\gcd(n,q)=1$. Assume that the order of $q$ modulo $n$ is equal to $m$. Let $\alpha$ be a primitive element of the finite field $\mathrm{GF}(q^m)$. Then $\beta=\alpha^{\frac{q^m-1}{n}}$ is a primitive $n$-th root of unity.

In this paper, we define
\begin{eqnarray*}
&N_1=\{n_1,2n_1,\cdots,(n_2-1)n_1\},&N_2=\{n_2,2n_2,\cdots,(n_1-1)n_2\},R=\{0\},\\
&C_0=R\cup N_2\cup D_3\cup D_4\cup D_5,&C_1=N_1\cup D_0\cup D_1\cup D_2.
\end{eqnarray*}
The binary Whiteman generalized cyclotomic sequence of order 6 $\bm{s^n}=(s_i)_{i=0}^{n-1}$ is defined by
$$
s_i=\left\{
\begin{array}{lll}
0, & \mbox{$\mathrm{if} \ i\in C_0$,}\\
1, & \mbox{$\mathrm{if} \ i\in C_1$.}
\end{array}
\right.
$$
Define the polynomials
\begin{equation}\label{eq4}
S(x)=\sum_{i\in C_1}x^i=\left(\sum_{i\in N_1}+\sum_{i\in D_0}+\sum_{i\in D_1}+\sum_{i\in D_2}\right)x^i,
\end{equation}
\begin{equation}\label{eq5}
T(x)=\left(\sum_{i\in N_1}+\sum_{i\in D_1}+\sum_{i\in D_2}+\sum_{i\in D_3}\right)x^i,
\end{equation}
\begin{equation}\label{eq6}
M(x)=\left(\sum_{i\in N_1}+\sum_{i\in D_2}+\sum_{i\in D_3}+\sum_{i\in D_4}\right)x^i,
\end{equation}
where $S(x)$, $T(x)$ and $M(x)\in \mathrm{GF}(q)[x]$. Then the polynomial $S(x)$ in Equation (\ref{eq4}) is exactly the generator polynomial of the sequence $\bm{s^n}$ \cite{Ref6}.

\section{Generator polynomials of cyclic codes}
Our main aim in this section is to find the generator polynomial $g(x)$ of the cyclic code $\bm{C_s}$ defined by the sequence $\bm{s^n}$, where $S(x)$ is the same as that in Equation (\ref{eq4}). Hence, we need only to find $a's$ such that $S(\beta^a)=0$ since $\beta$ is a primitive $n$-th root of unity, where $0\leq a \leq {n-1}$. To this end, we need the following Lemmas \ref{lem1}-\ref{lem5}.

\begin{lemma} \label{lem1}\cite{Ref22}
For any $r\in D_i$, we have $rD_j=D_{i+j(\mathrm{mod}\ 6)}$, where $rD_j=\{rd|d\in D_j\}.$
\end{lemma}

\begin{lemma}\label{lem2}
(I)$\displaystyle\sum\limits_{i\in N_1}\beta^i=\displaystyle\sum\limits_{i\in N_2}\beta^i=-1$; (II)$\displaystyle\sum\limits_{i\in Z_n^*}\beta^i=1$.
\end{lemma}
\noindent\textbf{Proof.} (I) can be proved directly in \cite{Ref23}. Now we will explain the result in (II). Since $$\beta^n-1=(\beta-1)\sum\limits_{i=0}^{n-1}\beta^i=0$$
and $\beta-1\neq 0$, we know that
$$\sum_{i=0}^{n-1}\beta^i=1+\sum_{i\in N_1}\beta^i+\sum_{i\in N_2}\beta^i+\sum_{i\in Z_n^*}\beta^i=0.$$
By (I), we obtain that
$$\sum_{i\in Z_n^*}\beta^i=\sum_{i\in \cup_{j=0}^5D_j}\beta^i=1+1-1=1.$$
This completes the proof of this lemma.$\square$

Note also that
$$
S(\beta^0)=S(1)=\displaystyle\frac{(n_1+1)(n_2-1)}{2}(\mathrm{mod}\ p).
$$

\begin{lemma}\label{lem3}
For $0\leq j\leq 5$, we have
$$
\sum_{i\in D_j}\beta^{ai}=\left\{
\begin{array}{ll}
-\displaystyle\frac{n_1-1}{6}(\bmod\ p), & \mbox{$\mathrm{if} \ a\in N_1$,}\vspace{1mm}\\
-\displaystyle\frac{n_2-1}{6}(\bmod\ p), & \mbox{$\mathrm{if} \ a\in N_2$.}
\end{array}
\right.
$$
\end{lemma}
\noindent\textbf{Proof.} Note that for $0\leq j\leq 5$, it holds that
\begin{eqnarray*}
D_j\bmod n_1 &=& \{g^sx^j\bmod n_1:s=0,1,\cdots,\displaystyle\frac{(n_1-1)(n_2-1)}{6}-1\}\\
             &=& \{g^{s+j}\bmod n_1:s=0,1,\cdots,\displaystyle\frac{(n_1-1)(n_2-1)}{6}-1\}\\
             &=& \displaystyle\frac{n_2-1}{6}*\{1,2,\cdots,n_1-1\},
\end{eqnarray*}
where $\displaystyle\frac{n_2-1}{6}$ is the multiplicity of each element in the set $\{1,2,\cdots,n_1-1\}$. In the similar way, we can prove that
$$D_j\bmod n_2=\displaystyle\frac{n_1-1}{6}*\{1,2,\cdots,n_2-1\}.$$
Suppose that $a\in N_1$. From Lemma \ref{lem2} (I), it holds that
$$\sum_{i\in D_j}\beta^i=\left(\displaystyle\frac{n_1-1}{6}\right)\sum_{i\in N_1}\beta^i=-\left(\displaystyle\frac{n_1-1}{6}\bmod p\right).$$
For $a\in N_2$, we can get the result by similar argument.$\square$
\begin{lemma}\label{lem4}
For $a\in Z_n$, we have
\begin{eqnarray*}
S(\beta^a)=T(\beta^a)=M(\beta^a)=\left\{
\begin{array}{lll}
-\displaystyle\frac{n_1+1}{2}(\bmod\ p),  & \mbox{$\mathrm{if} \ a \in N_1$,}\vspace{1mm}\\
\displaystyle\frac{n_2-1}{2}(\bmod\ p), & \mbox{$\mathrm{if} \ a \in N_2$,}\vspace{1mm}\\
\end{array}
\right.
\end{eqnarray*}
\begin{eqnarray*}
S(\beta^a)=\left\{
\begin{array}{lll}
S(\beta), & \mbox{$\mathrm{if} \ a \in D_0$,}\vspace{1mm}\\
T(\beta), & \mbox{$\mathrm{if} \ a \in D_1$,}\vspace{1mm}\\
M(\beta), & \mbox{$\mathrm{if} \ a \in D_2$,}\vspace{1mm}\\
-(S(\beta)+1), & \mbox{$\mathrm{if} \ a \in D_3$,}\vspace{1mm}\\
-(T(\beta)+1), & \mbox{$\mathrm{if} \ a \in D_4$,}\vspace{1mm}\\
-(M(\beta)+1), & \mbox{$\mathrm{if} \ a \in D_5$.}
\end{array}
\right.
\end{eqnarray*}
\begin{eqnarray*}
T(\beta^a)=\left\{
\begin{array}{lll}
T(\beta), & \mbox{$\mathrm{if} \ a \in D_0$,}\vspace{1mm}\\
M(\beta), & \mbox{$\mathrm{if} \ a \in D_1$,}\vspace{1mm}\\
-(S(\beta)+1), & \mbox{$\mathrm{if} \ a \in D_2$,}\vspace{1mm}\\
-(T(\beta)+1), & \mbox{$\mathrm{if} \ a \in D_3$,}\vspace{1mm}\\
-(M(\beta)+1), & \mbox{$\mathrm{if} \ a \in D_4$,}\vspace{1mm}\\
S(\beta), & \mbox{$\mathrm{if} \ a \in D_5$.}
\end{array}
\right.
\end{eqnarray*}
And
\begin{eqnarray*}
M(\beta^a)=\left\{
\begin{array}{lll}
M(\beta),      & \mbox{$\mathrm{if} \ a \in D_0$,}\vspace{1mm}\\
-(S(\beta)+1), & \mbox{$\mathrm{if} \ a \in D_1$,}\vspace{1mm}\\
-(T(\beta)+1), & \mbox{$\mathrm{if} \ a \in D_2$,}\vspace{1mm}\\
-(M(\beta)+1), & \mbox{$\mathrm{if} \ a \in D_3$,}\vspace{1mm}\\
S(\beta),      & \mbox{$\mathrm{if} \ a \in D_4$,}\vspace{1mm}\\
T(\beta),      & \mbox{$\mathrm{if} \ a \in D_5$.}
\end{array}
\right.
\end{eqnarray*}
\end{lemma}
\noindent\textbf{Proof.} For the case $aq\in N_1$, we have $aN_1\bmod n=\{a\omega\bmod  n:\omega\in N_1\}=N_1$ since $\gcd(n_1,n_2)=1$.
From Lemmas \ref{lem2} and \ref{lem3}, it follows that
\begin{eqnarray*}
S(\beta^a) &=& \left(\sum_{i\in N_1}+\sum_{i\in D_0}+\sum_{i\in D_1}+\sum_{i\in D_2}\right)\beta^{ai}\\
           &=& \sum_{i\in N_1}\beta^i+\sum_{i\in D_0}\beta^{ai}+\sum_{i\in D_1}\beta^{ai}+\sum_{i\in D_2}\beta^{ai}\\
           &=& -1-\left(\displaystyle\frac{n_1-1}{6}\bmod p\right)-\left(\displaystyle\frac{n_1-1}{6}\bmod p\right) -\left(\displaystyle\frac{n_1-1}{6}\bmod p\right)\\
           &=& -\displaystyle\frac{n_1+1}{2}\bmod p.
\end{eqnarray*}
Similarly, $T(\beta^a)=-\displaystyle\frac{n_1+1}{2}\bmod p$ and $M(\beta^a)=-\displaystyle\frac{n_1+1}{2}\bmod p$ when $a\in N_1$.

For $a\in N_2$, we have $aN_1\mathrm{mod}\ n=\{0\}.$ By Lemma \ref{lem3} again, we can get
\begin{eqnarray*}
S(\beta^a) &=& \left(\sum_{i\in N_1}+\sum_{i\in D_0}+\sum_{i\in D_1}+\sum_{i\in D_2}\right)\beta^{ai}\\
           &=& \sum_{i\in N_1}1+\sum_{i\in D_0}\beta^{ai}+\sum_{i\in D_1}\beta^{ai}+\sum_{i\in D_2}\beta^{ai}\\
           &=& n_2-1-\left(\displaystyle\frac{n_2-1}{6}\bmod p\right)-\left(\displaystyle\frac{n_2-1}{6}\bmod p\right)-\left(\displaystyle\frac{n_2-1}{6}\bmod p\right)\\
           &=& \displaystyle\frac{n_2-1}{2}\bmod p.
\end{eqnarray*}
Similarly, we can prove $T(\beta^a)=\displaystyle\frac{n_2-1}{2}\bmod p$ and $M(\beta^a)=\displaystyle\frac{n_2-1}{2}\bmod p$ when $a\in N_2$.

For the case that $a\in D_0$, we have $aD_j=D_{j(\bmod 6)}$, $aN_1\bmod n=N_1$. By Lemma \ref{lem2} and Equation (\ref{eq4}), we get
\begin{eqnarray*}
S(\beta^a) &=& \left(\sum_{i\in N_1}+\sum_{i\in D_0}+\sum_{i\in D_1}+\sum_{i\in D_2}\right)\beta^{ai}\\
           &=& \left(\sum_{i\in N_1}+\sum_{i\in D_0}+\sum_{i\in D_1}+\sum_{i\in D_2}\right)\beta^i\\
           &=& S(\beta).
\end{eqnarray*}
Similarly, we can obtain $T(\beta^a)=T(\beta)$ and $M(\beta^a)=M(\beta)$ when $a\in D_0$.

For the case $a\in D_1$,  we have $aD_j=D_{j+1(\bmod 6)}$, $aN_1\bmod n=N_1$. By Lemma \ref{lem2} and Equation (\ref{eq5}), we get
\begin{eqnarray*}
S(\beta^a) &=& \left(\sum_{i\in N_1}+\sum_{i\in D_0}+\sum_{i\in D_1}+\sum_{i\in D_2}\right)\beta^{ai}\\
           &=& \left(\sum_{i\in N_1}+\sum_{i\in D_1}+\sum_{i\in D_2}+\sum_{i\in D_3}\right)\beta^i\\
           &=& T(\beta).
\end{eqnarray*}
When $a\in D_1$, $T(\beta^a)=M(\beta)$ and $M(\beta^a)=-(S(\beta)+1)$ can be proved in the same way.

For $a\in D_2$, $aD_j=D_{j+2(\bmod 6)}$, $aN_1\bmod n=N_1$. it derives from Lemma \ref{lem2} and Equation (\ref{eq6}) that
\begin{eqnarray*}
S(\beta^a) &=& \left(\sum_{i\in N_1}+\sum_{i\in D_0}+\sum_{i\in D_1}+\sum_{i\in D_2}\right)\beta^{ai}\\
           &=& \left(\sum_{i\in N_1}+\sum_{i\in D_2}+\sum_{i\in D_3}+\sum_{i\in D_4}\right)\beta^i\\
           &=& M(\beta).
\end{eqnarray*}
A similar proof can deduce $T(\beta^a)=-(S(\beta)+1)$ and $M(\beta^a)=-(T(\beta)+1)$ when $a\in D_2$.

For $a\in D_3$, $aD_j=D_{j+3(\bmod 6)}$, $aN_1\bmod n=N_1$. By Lemma \ref{lem2} and Equation (\ref{eq4}), it holds that
\begin{eqnarray*}
S(\beta^a) &=& \left(\sum_{i\in N_1}+\sum_{i\in D_0}+\sum_{i\in D_1}+\sum_{i\in D_2}\right)\beta^{ai}\\
           &=& \left(\sum_{i\in N_1}+\sum_{i\in D_3}+\sum_{i\in D_4}+\sum_{i\in D_5}\right)\beta^i\\
           &=& \sum_{i\in N_1}\beta^i+1-\left(\sum_{i\in D_0}+\sum_{i\in D_1}+\sum_{i\in D_2}\right)\beta^i\\
           &=& 2\sum_{i\in N_1}\beta^i+1-\left(\sum_{i\in N_1}+\sum_{i\in D_0}+\sum_{i\in D_1}+\sum_{i\in D_2}\right)\beta^i\\
           &=& -(S(\beta)+1).
\end{eqnarray*}
Similarly, we have $T(\beta^a)=-(T(\beta)+1)$ and $M(\beta^a)=-(M(\beta)+1)$ when $a\in D_3$.

For $a\in D_4$, $aD_j=D_{j+4(\bmod 6)}$, $aN_1\bmod n=N_1$. By Lemma \ref{lem2} and Equation (\ref{eq5}), we get
\begin{eqnarray*}
S(\beta^a) &=& \left(\sum_{i\in N_1}+\sum_{i\in D_0}+\sum_{i\in D_1}+\sum_{i\in D_2}\right)\beta^{ai}\\
           &=& \left(\sum_{i\in N_1}+\sum_{i\in D_4}+\sum_{i\in D_5}+\sum_{i\in D_0}\right)\beta^i\\
           &=& \sum_{i\in N_1}\beta^i+1-\left(\sum_{i\in D_0}+\sum_{i\in D_1}+\sum_{i\in D_2}\right)\beta^i\\
           &=& 2\sum_{i\in N_1}\beta^i+1-\left(\sum_{i\in N_1}+\sum_{i\in D_1}+\sum_{i\in D_2}+\sum_{i\in D_3}\right)\beta^i\\
           &=& -(T(\beta)+1).
\end{eqnarray*}
when $a\in D_4$, $T(\beta^a)=-(M(\beta)+1)$ and $M(\beta^a)=S(\beta)$ can be proved similarly.

For $a\in D_5$, $aD_j=D_{j+5(\bmod 6)}$, $aN_1\bmod n=N_1$. By Lemma \ref{lem2} and Equation (\ref{eq6}), we have
\begin{eqnarray*}
S(\beta^a) &=& \left(\sum_{i\in N_1}+\sum_{i\in D_0}+\sum_{i\in D_1}+\sum_{i\in D_2}\right)\beta^{ai}\\
           &=& \left(\sum_{i\in N_1}+\sum_{i\in D_5}+\sum_{i\in D_0}+\sum_{i\in D_1}\right)\beta^i\\
           &=& \sum_{i\in N_1}\beta^i+1-\left(\sum_{i\in D_2}+\sum_{i\in D_3}+\sum_{i\in D_4}\right)\beta^i\\
           &=& 2\sum_{i\in N_1}\beta^i+1-\left(\sum_{i\in N_1}+\sum_{i\in D_2}+\sum_{i\in D_3}+\sum_{i\in D_4}\right)\beta^i\\
           &=& -(M(\beta)+1).
\end{eqnarray*}
A similar proof will deduce $T(\beta^a)=S(\beta)$ and $M(\beta^a)=T(\beta)$ when $a\in D_5$.$\square$

\begin{lemma}\label{lem5}
Let the symbol be defined as before.\\
(I) If $q\in D_1\cup D_3\cup D_5$, we have $S(\beta)\notin \{-1,0\}$, $T(\beta)\notin \{-1,0\}$ and $M(\beta)\notin \{-1,0\}$.\\
(II) If $q\in D_0$, we have $[S(\beta)]^q=S(\beta)$, $[T(\beta)]^q=T(\beta)$ and $[M(\beta)]^q=M(\beta)$.\\
(III) If $q\in D_2\cup D_4$, we have $[S(\beta)]^{q^3}=S(\beta)$, $[T(\beta)]^{q^3}=T(\beta)$ and $[M(\beta)]^{q^3}=M(\beta)$.
\end{lemma}
\noindent\textbf{Proof.} (I) If $q\in D_1$, then by Lemma \ref{lem4}, it holds that
$$[S(\beta)]^{q^3}=[S(\beta^q)]^{q^2}=[T(\beta^q)]^q=[M(\beta)]^q=-(S(\beta)+1).$$
Namely,
$$[S(\beta)]^{q^3}+S(\beta)+1=0.$$
It is easy to check that none of 0 and -1 is a solution of the equation above.

When $q\in D_3\cup D_5$, the same result can be obtained.

Similarly, it holds that $T(\beta)\notin \{-1,0\}$ and $M(\beta)\notin \{-1,0\}$ when $q\in D_1\cup D_3\cup D_5$.

(II) If $q\in D_0$, then by Lemma \ref{lem4}, we have
$$[S(\beta)]^q=S(\beta), [T(\beta)]^q=T(\beta), [M(\beta)]^q=M(\beta).$$
(III) If $q\in D_2$, then by Lemma \ref{lem4},
$$
[S(\beta)]^{q^3}=[S(\beta)]^{q^2}=[M(\beta^q)]^q=[-T(\beta)-1]^q=[(S(\beta)+1)-1]=S(\beta).
$$
When $q\in D_4$, we can get $[S(\beta)]^{q^3}=S(\beta)$ in the same way. When $q\in D_2\cup D_4$, we can prove that $[T(\beta)]^{q^3}=T(\beta)$ and $[M(\beta)]^{q^3}=M(\beta)$ similarly.$\square$

Let $\beta$ be the same as before. Among the $n_1n_2$ $n_1n_2$-th roots of unity $\beta^i$, where $0\leq i \leq n_1n_2-1$, the $n_2$ elements $\beta^i$, $i\in N_1\cup R$, are $n_2$-th roots of unity, the $n_1$ elements $\beta^i$, $i\in N_2\cup R$, are $n_1$-th roots of unity. Hence,
$$x^{n_1}-1=\prod_{i\in N_2\cup R}(x-\beta^i),\ \ x^{n_2}-1=\prod_{i\in N_1\cup R}(x-\beta^i).$$
Let
$$d(x)=\prod_{i\in Z_n^*}(x-\beta^i).$$
It follows that
$$x^n-1=\prod_{i=0}^{n_1n_2-1}(x-\beta^i)=\displaystyle\frac{(x^{n_1}-1)(x^{n_2}-1)}{x-1}d(x),$$
where $d(x)\in \mathrm{GF}(q)[x].$

Let $$\Omega_i=\displaystyle\frac{n_i+(-1)^{i-1}}{2}(\bmod\ p)$$ and $$\Omega=S(\beta^0)=\displaystyle\frac{(n_1+1)(n_2-1)}{2}(\bmod\ p).$$ Then from Lemma \ref{lem5}, we have the following theorem.

\begin{theorem}\label{thm1}
Assume that $q\in D_1\cup D_3\cup D_5$. Then
\begin{eqnarray*}
g(x)=\left\{
\begin{array}{llll}
x^n-1,  & \Omega_1\neq 0,\Omega_2\neq 0,\Omega\neq 0,\vspace{1mm}\\
\displaystyle\frac{x^n-1}{x-1}, & \Omega_1\neq 0,\Omega_2\neq 0,\Omega=0,\vspace{1mm}\\
\displaystyle\frac{x^n-1}{x^{n_2}-1}, & \Omega_1=0,\Omega_2\neq 0,\Omega=0,\vspace{1mm}\\
\displaystyle\frac{x^n-1}{x^{n_1}-1}, & \Omega_1\neq 0,\Omega_2=0,\Omega=0,\vspace{1mm}\\
d(x), & \Omega_1=\Omega_2=\Omega=0.
\end{array}
\right.
\end{eqnarray*}
In this case, the cyclic code $\bm{C_s}$ over $\mathrm{GF}(q)$ defined by the Whiteman generalized cyclotomic sequence $\bm{s^n}$ of order 6 has generator polynomial $g(x)$ as above.
\end{theorem}
\noindent\textbf{Proof.} Note that $\Omega_i=0$ ($i$=0,1) results in $\Omega=0$ and that both $\Omega_1=0$ and  $\Omega_2=0$ result in $\Omega=0$. By Lemma \ref{lem5}, we know that if $q\in D_1\cup D_3\cup D_5$, then $S(\beta)\notin \{-1,0\}$.

\textbf{Case 1:} $\Omega_1\neq 0,\Omega_2\neq 0,\Omega\neq 0$, by Lemma \ref{lem4}, it holds that
$$\gcd(S(x),x^n-1)=1,$$
then by Equation (\ref{eq2}),
$$g(x)=\displaystyle\frac{x^n-1}{\gcd(S(x),x^n-1)}=x^n-1.$$
\textbf{Case 2:} $\Omega_1\neq 0,\Omega_2\neq 0,\Omega=0$, by Lemma \ref{lem4}, it follows that
$$\gcd(S(x),x^n-1)=x-1,$$
then by Equation (\ref{eq2}),
$$g(x)=\displaystyle\frac{x^n-1}{\gcd(S(x),x^n-1)}=\displaystyle\frac{x^n-1}{x-1}.$$
\textbf{Case 3:} $\Omega_1=0,\Omega_2\neq 0,\Omega=0$, by Lemma \ref{lem4}, we can get
$$\gcd(S(x),x^n-1)=x^{n_2}-1,$$
then by Equation (\ref{eq2}),
$$g(x)=\displaystyle\frac{x^n-1}{\gcd(S(x),x^n-1)}=\displaystyle\frac{x^n-1}{x^{n_2}-1}.$$
\textbf{Case 4:} $\Omega_1\neq 0,\Omega_2=0,\Omega=0$, by Lemma \ref{lem4}, we have
$$\gcd(S(x),x^n-1)=x^{n_1}-1,$$
then by Equation (\ref{eq2}),
$$g(x)=\displaystyle\frac{x^n-1}{\gcd(S(x),x^n-1)}=\displaystyle\frac{x^n-1}{x^{n_1}-1}.$$
\textbf{Case 5:} $\Omega_1=\Omega_2=\Omega=0$, by Lemma \ref{lem4}, we can obtain
$$\gcd(S(x),x^n-1)=\displaystyle\frac{(x^{n_1}-1)(x^{n_2}-1)}{x-1},$$
then by Equation (\ref{eq2}),
$$g(x)=\displaystyle\frac{x^n-1}{\gcd(S(x),x^n-1)}=\displaystyle\frac{(x^n-1)(x-1)}{(x^{n_1}-1)(x^{n_2}-1)}.$$
This completes the proof of this theorem.$\square$

\textbf{Example 1} Let $p=2$, $n_1=7, n_2=13$. Then $q=2, n=91, \Omega_1=\Omega_2=\Omega=0$, and $\bm{C_s}$ is a [91, 72, 4] cyclic code over $\mathrm{GF}(q)$ with generator polynomial $x^{19}+x^{18}+x^{17}+x^{16}+x^{15}+x^{14}+x^{13}+x^6+x^5+x^4+x^3+x^2+x+1$.

\textbf{Example 2} Let $p=2$, $n_1=13, n_2=7$. Then $q=2, n=91, \Omega_1=\Omega_2=1, \Omega=0$, and $\bm{C_s}$ is a [91, 1, 91] cyclic code over $\mathrm{GF}(q)$ with generator polynomial $\displaystyle\frac{x^{91}-1}{x-1}$.

Define
$$d_a(x)=\prod_{i\in D_a}(x-\beta^i), a=0, 1, \cdots, 5.$$

In case $q\in D_0$, by Lemma 1, for all $a=0, 1, \cdots, 5,$
$$
[d_a(x)]^q =\prod_{i\in D_a}(x^q-\beta^{qi})=\prod_{k\in qD_a}(x^q-\beta^k)=\prod_{k\in D_a}(x^q-\beta^k)=d_a(x^q).
$$
Hence, $d_a(x)\in \mathrm{GF}(q)[x]$, $j=0, 1, \cdots, 5.$ Thus, in case $q\in D_0$ we get
\begin{equation}\label{eq7}
x^n-1=\displaystyle\frac{(x^{n_1}-1)(x^{n_2}-1)}{x-1}\prod_{i=0}^5d_i(x).
\end{equation}
If $q\in D_0$, by Lemma 5, we know that $S(x)$, $T(x)$, $M(x)\in \mathrm{GF}(q)[x]$. And it is possible for $S(\beta)\in \{0, -1\}$, $T(\beta)\in \{0, -1\}$ and $M(\beta)\in \{0, -1\}$. Then from the definition as above, we have the following theorem.

\begin{theorem}
Assume that $q\in D_0\cup D_2\cup D_4$. Then\\
(I) If one of $S(\beta)$, $T(\beta)$, $M(\beta)$ is in $\{0, -1\}$, then
\begin{eqnarray*}
g(x)=\left\{
\begin{array}{lll}
\displaystyle\frac{x^n-1}{d_m(x)},& \Omega_1\neq 0,\Omega_2\neq 0,\Omega\neq 0,\vspace{1mm}\\
\displaystyle\frac{x^n-1}{(x-1)d_m(x)}, & \Omega_1\neq 0,\Omega_2\neq 0,\Omega=0,\vspace{1mm}\\
\displaystyle\frac{x^n-1}{(x^{n_2}-1)d_m(x)},& \Omega_1=0,\Omega_2\neq 0,\Omega=0,\vspace{1mm}\\
\displaystyle\frac{x^n-1}{(x^{n_1}-1)d_m(x)},& \Omega_1\neq 0,\Omega_2=0,\Omega=0,\vspace{1mm}\\
\displaystyle\frac{d(x)}{d_m(x)},& \Omega_1=\Omega_2=\Omega=0,\vspace{1mm}
\end{array}
\right.
\end{eqnarray*}
where
\begin{eqnarray*}
m=\left\{
\begin{array}{lll}
0,\quad S(\beta)=0\vspace{1mm},\\
3,\quad S(\beta)=-1\vspace{1mm},
\end{array}
\right.
m=\left\{
\begin{array}{lll}
1,\quad T(\beta)=0\vspace{1mm},\\
4,\quad T(\beta)=-1\vspace{1mm}.\\
\end{array}
\right.
m=\left\{
\begin{array}{lll}
2,\quad M(\beta)=0\vspace{1mm},\\
5,\quad M(\beta)=-1\vspace{1mm}.
\end{array}
\right.
\end{eqnarray*}
(II) If two of $S(\beta)$, $T(\beta)$, $M(\beta)$ are in $\{0, -1\}$, the specific generator polynomials are listed as follows.

(i) if $S(\beta), T(\beta)\in \{0,-1\}$, then
\begin{eqnarray*}
g(x)=\left\{
\begin{array}{lll}
\displaystyle\frac{x^n-1}{d_m(x)d_s(x)},& \Omega_1\neq 0,\Omega_2\neq 0,\Omega\neq 0,\vspace{1mm}\\
\displaystyle\frac{x^n-1}{(x-1)d_m(x)d_s(x)},& \Omega_1\neq 0,\Omega_2\neq 0,\Omega=0,\vspace{1mm}\\
\displaystyle\frac{x^n-1}{(x^{n_2}-1)d_m(x)d_s(x)},& \Omega_1=0,\Omega_2\neq 0,\Omega=0,\vspace{1mm}\\
\displaystyle\frac{x^n-1}{(x^{n_1}-1)d_m(x)d_s(x)},& \Omega_1\neq 0,\Omega_2=0,\Omega=0,\vspace{1mm}\\
\displaystyle\frac{d(x)}{d_m(x)d_s(x)},& \Omega_1=\Omega_2=\Omega=0,\vspace{1mm}
\end{array}
\right.
\end{eqnarray*}
where
\begin{eqnarray*}
m=\left\{
\begin{array}{lll}
0,\quad S(\beta)=0\vspace{1mm},\\
3,\quad S(\beta)=-1\vspace{1mm},
\end{array}
\right.
 \ \ s=\left\{
\begin{array}{lll}
1,\quad T(\beta)=0\vspace{1mm},\\
4,\quad T(\beta)=-1\vspace{1mm}.\\
\end{array}
\right.
\end{eqnarray*}

(ii) if $S(\beta), M(\beta)\in \{0,-1\}$, then
\begin{eqnarray*}
g(x)=\left\{
\begin{array}{lll}
\displaystyle\frac{x^n-1}{d_m(x)d_t(x)},& \Omega_1\neq 0,\Omega_2\neq 0,\Omega\neq 0,\vspace{1mm}\\
\displaystyle\frac{x^n-1}{(x-1)d_m(x)d_t(x)},& \Omega_1\neq 0,\Omega_2\neq 0,\Omega=0,\vspace{1mm}\\
\displaystyle\frac{x^n-1}{(x^{n_2}-1)d_m(x)d_t(x)},& \Omega_1=0,\Omega_2\neq 0,\Omega=0,\vspace{1mm}\\
\displaystyle\frac{x^n-1}{(x^{n_1}-1)d_m(x)d_t(x)},& \Omega_1\neq 0,\Omega_2=0,\Omega=0,\vspace{1mm}\\
\displaystyle\frac{d(x)}{d_m(x)d_s(x)},& \Omega_1=\Omega_2=\Omega=0,\vspace{1mm}
\end{array}
\right.
\end{eqnarray*}
where
\begin{eqnarray*}
m=\left\{
\begin{array}{lll}
0,\quad S(\beta)=0\vspace{1mm},\\
3,\quad S(\beta)=-1\vspace{1mm},
\end{array}
\right.
\ \ t=\left\{
\begin{array}{lll}
2,\quad M(\beta)=0\vspace{1mm},\\
5,\quad M(\beta)=-1\vspace{1mm}.
\end{array}
\right.
\end{eqnarray*}

(iii) if $T(\beta), M(\beta)\in \{0,-1\}$, then
\begin{eqnarray*}
g(x)=\left\{
\begin{array}{lll}
\displaystyle\frac{x^n-1}{d_s(x)d_t(x)},& \Omega_1\neq 0,\Omega_2\neq 0,\Omega\neq 0,\vspace{1mm}\\
\displaystyle\frac{x^n-1}{(x-1)d_s(x)d_t(x)},& \Omega_1\neq 0,\Omega_2\neq 0,\Omega\neq 0,\vspace{1mm}\\
\displaystyle\frac{x^n-1}{(x^{n_2}-1)d_s(x)d_t(x)},& \Omega_1=0,\Omega_2\neq 0,\Omega=0,\vspace{1mm}\\
\displaystyle\frac{x^n-1}{(x^{n_1}-1)d_s(x)d_t(x)},& \Omega_1\neq 0,\Omega_2=0,\Omega=0,\vspace{1mm}\\
\displaystyle\frac{d(x)}{d_s(x)d_t(x)},& \Omega_1=\Omega_2=\Omega=0,\vspace{1mm}
\end{array}
\right.
\end{eqnarray*}
where
\begin{eqnarray*}
s=\left\{
\begin{array}{lll}
1,\quad T(\beta)=0\vspace{1mm},\\
4,\quad T(\beta)=-1\vspace{1mm},\\
\end{array}
\right.
\ \
t=\left\{
\begin{array}{lll}
2,\quad M(\beta)=0\vspace{1mm},\\
5,\quad M(\beta)=-1\vspace{1mm}.
\end{array}
\right.
\end{eqnarray*}
(III) If $S(\beta), T(\beta), M(\beta)\in \{0,-1\}$, then
\begin{eqnarray*}
g(x)=\left\{
\begin{array}{lll}
\displaystyle\frac{x^n-1}{d_m(x)d_s(x)d_t(x)},& \Omega_1\neq 0,\Omega_2\neq 0,\Omega\neq 0,\vspace{1mm}\\
\displaystyle\frac{x^n-1}{(x-1)d_m(x)d_s(x)d_t(x)},& \Omega_1\neq 0,\Omega_2\neq 0,\Omega= 0,\vspace{1mm}\\
\displaystyle\frac{x^n-1}{(x^{n_2}-1)d_m(x)d_s(x)d_t(x)},& \Omega_1=0,\Omega_2\neq 0,\Omega=0,\vspace{1mm}\\
\displaystyle\frac{x^n-1}{(x^{n_1}-1)d_m(x)d_s(x)d_t(x)},& \Omega_1\neq 0,\Omega_2=0,\Omega=0,\vspace{1mm}\\
\displaystyle\frac{d(x)}{d_m(x)d_s(x)d_t(x)},& \Omega_1=\Omega_2=\Omega=0,\vspace{1mm}
\end{array}
\right.
\end{eqnarray*}
where
\begin{eqnarray*}
m=\left\{
\begin{array}{lll}
0,\quad S(\beta)=0\vspace{1mm},\\
3,\quad S(\beta)=-1\vspace{1mm},
\end{array}
\right.
s=\left\{
\begin{array}{lll}
1,\quad T(\beta)=0\vspace{1mm},\\
4,\quad T(\beta)=-1\vspace{1mm},
\end{array}
\right.
t=\left\{
\begin{array}{lll}
2,\quad M(\beta)=0\vspace{1mm},\\
5,\quad M(\beta)=-1\vspace{1mm}.
\end{array}
\right.
\end{eqnarray*}
In the cases above, the cyclic code $\bm{C_s}$ over $\mathrm{GF}(q)$ defined by the Whiteman generalized cyclotomic sequence $\bm{s^n}$ of order 6 has generator polynomial $g(x)$ as above correspondingly.
\end{theorem}
\noindent\textbf{Proof.} Note that $\Omega_i=0$ results in $\Omega=0$ and that both $\Omega_1=0$ and  $\Omega_2=0$ lead to $\Omega=0$. By Lemma 5, we know that if $q\in D_0\cup D_2\cup D_4$, then it is possible for $S(\beta)\in \{0, -1\}$, $T(\beta)\in \{0, -1\}$ and $M(\beta)\in \{0, -1\}$.

In the following, we only give the proof of the case that $S(\beta)\in \{0, -1\}$, and the other cases can be proved similarly.

\textbf{Case 1:} $\Omega_1\neq 0$, $\Omega_2\neq 0$, $\Omega\neq 0, S(\beta)=0$, by Lemma \ref{lem4}, we can obtain
$$\gcd(S(x),x^n-1)=d_0(x),$$
then by Equations (\ref{eq2}) and (\ref{eq7}),
$$g(x)=\displaystyle\frac{x^n-1}{\gcd(S(x),x^n-1)}=\displaystyle\frac{x^n-1}{d_0(x)},$$
and it means that $m=0$.

\textbf{Case 2:} $\Omega_1\neq 0$, $\Omega_2\neq 0$, $\Omega\neq 0, S(\beta)=-1$, by Lemma \ref{lem4} again, we have
$$\gcd(S(x),x^n-1)=d_3(x),$$
so we can obtain the generator polynomial by Equations (\ref{eq2}) and (\ref{eq7}),
$$g(x)=\displaystyle\frac{x^n-1}{\gcd(S(x),x^n-1)}=\displaystyle\frac{x^n-1}{d_3(x)},$$
which implies that $m=3$.

\textbf{Case 3:} $\Omega_1\neq 0$, $\Omega_2\neq 0$, $\Omega=0, S(\beta)=0$, by Lemma \ref{lem4}, it holds that
$$\gcd(S(x),x^n-1)=(x-1)d_0(x),$$
by Equations (\ref{eq2}) and (\ref{eq7}), the corresponding generator polynomial is as follows,
$$g(x)=\displaystyle\frac{x^n-1}{\gcd(S(x),x^n-1)}=\displaystyle\frac{x^n-1}{(x-1)d_0(x)},$$
which means that $m=0$.

\textbf{Case 4:} $\Omega_1\neq 0$, $\Omega_2\neq 0$, $\Omega=0, S(\beta)=-1$, by Lemma \ref{lem4}, it suffices to get
$$\gcd(S(x),x^n-1)=(x-1)d_3(x),$$
then by Equations (\ref{eq2}) and (\ref{eq7}), the corresponding generator polynomial is given by
$$g(x)=\displaystyle\frac{x^n-1}{\gcd(S(x),x^n-1)}=\displaystyle\frac{x^n-1}{(x-1)d_3(x)},$$
and $m=3$ obviously.

\textbf{Case 5:} $\Omega_1=0$, $\Omega_2\neq 0$, $\Omega=0, S(\beta)=0$, by Lemma \ref{lem4}, we can get
$$\gcd(S(x),x^n-1)=(x^{n_2}-1)d_0(x),$$
then the generator polynomial can be calculated by Equations (\ref{eq2}) and (\ref{eq7}),
$$g(x)=\displaystyle\frac{x^n-1}{\gcd(S(x),x^n-1)}=\displaystyle\frac{x^n-1}{(x^{n_2}-1)d_0(x)},$$
and it implies that $m=0$.

\textbf{Case 6:} $\Omega_1=0$, $\Omega_2\neq 0$, $\Omega=0, S(\beta)=-1$, by Lemma \ref{lem4}, it holds that
$$\gcd(S(x),x^n-1)=(x^{n_2}-1)d_3(x),$$
then the generator polynomial follows from Equations (\ref{eq2}) and (\ref{eq7}),
$$g(x)=\displaystyle\frac{x^n-1}{\gcd(S(x),x^n-1)}=\displaystyle\frac{x^n-1}{(x^{n_2}-1)d_3(x)},$$
which means that $m=3$.

\textbf{Case 7:} $\Omega_1\neq 0$, $\Omega_2=0$, $\Omega=0, S(\beta)=0$, by Lemma \ref{lem4}, we can obtain
$$\gcd(S(x),x^n-1)=(x^{n_1}-1)d_0(x),$$
then by Equations (\ref{eq2}) and (\ref{eq7}), it suffices to have
$$g(x)=\displaystyle\frac{x^n-1}{\gcd(S(x),x^n-1)}=\displaystyle\frac{x^n-1}{(x^{n_1}-1)d_0(x)},$$
which implies that $m=0$.

\textbf{Case 8:} $\Omega_1\neq 0$, $\Omega_2=0$, $\Omega=0, S(\beta)=-1$, by Lemma \ref{lem4}, a similar proof will deduce
$$\gcd(S(x),x^n-1)=(x^{n_1}-1)d_3(x),$$
by Equations (\ref{eq2}) and (\ref{eq7}) again, we can get
$$g(x)=\displaystyle\frac{x^n-1}{\gcd(S(x),x^n-1)}=\displaystyle\frac{x^n-1}{(x^{n_1}-1)d_3(x)},$$
and $m=3$ obviously.

\textbf{Case 9:} $\Omega_1=\Omega_2=\Omega=0, S(\beta)=0$, by Lemma \ref{lem4}, we have
$$\gcd(S(x),x^n-1)=\displaystyle\frac{(x^{n_1}-1)(x^{n_2}-1)}{x-1}d_0(x),$$
by Equations (\ref{eq2}) and (\ref{eq7}) again, the generator polynomial can be given by
$$g(x)=\displaystyle\frac{x^n-1}{\gcd(S(x),x^n-1)}=\displaystyle\frac{d(x)}{d_0(x)},$$
and it means that $m=0$.

\textbf{Case 10:} $\Omega_1=\Omega_2=\Omega=0, S(\beta)=-1$, by Lemma \ref{lem4}, we can get
$$\gcd(S(x),x^n-1)=\displaystyle\frac{(x^{n_1}-1)(x^{n_2}-1)}{x-1}d_3(x),$$
then by Equations (\ref{eq2}) and (\ref{eq7})
$$g(x)=\displaystyle\frac{x^n-1}{\gcd(S(x),x^n-1)}=\displaystyle\frac{d(x)}{d_3(x)},$$
and $m=3$ obviously.

The rest results of this theorem can be proved similarly.$\square$

\noindent\textbf{Example 3} Let $q=2$, $n_1=13, n_2=19$. Then $p=2, n=247, \Omega_1=\Omega_2=1, \Omega=0, S(\beta)=T(\beta)=0, M(\beta)=1$, and $\bm{C_s}$ is a [247, 109] cyclic code over $\mathrm{GF}(q)$ with generator polynomial $\displaystyle\frac{x^{247}-1}{(x-1)d_0(x)d_1(x)d_5(x)}$.

\noindent\textbf{Example 4} Let $q=2$, $n_1=19, n_2=13$. Then $p=2, n=247, \Omega_1=\Omega_2=\Omega=0, S(\beta)=T(\beta)=M(\beta)=1$, and $\bm{C_s}$ is a [247, 139] cyclic code over $\mathrm{GF}(q)$ with generator polynomial $d_0(x)d_1(x)d_2(x)$.

\noindent\textbf{Example 5} Let $q=3$, $n_1=31, n_2=19$. Then $p=3, n=589, \Omega_1=0, \Omega_2=1, \Omega=0, S(\beta)=T(\beta)=-1, M(\beta)=0$, and $\bm{C_s}$ is a [589, 289] cyclic code over $\mathrm{GF}(q)$ with generator polynomial $\displaystyle\frac{x^{589}-1}{(x^{19}-1)d_2(x)d_3(x)d_4(x)}$.

\noindent\textbf{Example 6} Let $q=3$, $n_1=19, n_2=31$. Then $p=3, n=589, \Omega_1=0, \Omega_2=1, \Omega=0, S(\beta)=T(\beta)=M(\beta)=0$, and $\bm{C_s}$ is a [589, 301] cyclic code over $\mathrm{GF}(q)$ with generator polynomial $\displaystyle\frac{x^{589}-1}{(x^{31}-1)d_0(x)d_1(x)d_2(x)}$.

\section{The minimal distance of the cyclic codes}
In this section, we calculate the minimum distance of some cyclic codes and give lower bounds of the minimum distance for some other cyclic codes in Section 2. By the same argument as that in \cite{Ref16} and \cite{Ref18}, we can get the following two results immediately.
\begin{theorem}\label{thm3}
Let $\bm{C_i}$ denote the cyclic code over $\mathrm{GF}(q)$ generated by the polynomial $g_i(x)=\displaystyle\frac{x^n-1}{x^{n_i}-1}$. The $\bm{C_i}$ has parameters $[n$, $n_i$, $d_i]$, where $d_i=n_{i-(-1)^i}$ and $i=1,2$.
\end{theorem}

\noindent\textbf{Example 7} Let $q=2$ and $n_1=13,n_2=31$. Then the cyclic code over $\mathrm{GF}(q)$ with the generator polynomial $g(x)=\displaystyle\frac{x^n-1}{x^{n_1}-1}$ has parameters [403, 13, 31].
\begin{theorem}\label{thm4}
Let $\bm{C_{n_1,n_2}}$ denote the cyclic code over $\mathrm{GF}(q)$ generated by the polynomial $g_{n_1,n_2}(x)=\displaystyle\frac{(x^n-1)(x-1)}{(x^{n_1}-1)(x^{n_2}-1)}$. The $\bm{C_{n_1,n_2}}$ has parameters $[n$, $n_1+n_2-1$, $d_{n_1,n_2}]$, where $d_{n_1,n_2}=\min(n_1,n_2)$.
\end{theorem}

\noindent\textbf{Example 8} Let $q=2$ and $n_1=13,n_2=31$. Then the cyclic code over $\mathrm{GF}(q)$ with the generator polynomial $g(x)=\displaystyle\frac{(x^n-1)(x-1)}{(x^{n_1}-1)(x^{n_2}-1)}$ has parameters [403, 43, 13].

We also derive the following lower bounds of the minimum distances of other cyclic codes.

\begin{theorem}\label{thm5}
Suppose that $q \in D_0$. Let $\bm{C_i}$ and $d_i$ be defined as in Theorem \ref{thm3}. Let $\bm{C_{i,j}}$ denote the cyclic code over $\mathrm{GF}(q)$ generated by the polynomial $g_{i,j}(x)=\displaystyle\frac{x^n-1}{(x^{n_i}-1)d_j(x)}$. The $\bm{C_{i,j}}$ has parameters $[n$, $n_i+\displaystyle\frac{(n_1-1)(n_2-1)}{6}$, $d_{i,j}]$, where $d_{i,j}\geq \lceil \sqrt{n_{i-(-1)^i}} \rceil$, $i=1,2$ , and $0\leq j \leq5$.
\end{theorem}
\noindent\textbf{Proof.} Note that for $0\leq j \leq5$ and for any $r\in D_j$, we have $r^{-1}(\bmod n)\in D_{(6-i)\bmod 6}$. Let $i=1, j=0$ and $c(x)\in \mathrm{GF}(q)[x]/(x^n-1)$ be a codeword of Hamming weight $w$ in $\bm{C_{1,0}}$. Take $r\in D_1$, then $r^{-1}(\bmod n)\in D_5$ and $c(x^r)$ is a codeword of Hamming weight $w$ in $\bm{C_{1,5}}$, which implies that $d_{1,0}=d_{1,5}$. By taking $r\in D_j$, we can get $d_{1,0}=d_{1,6-j}$, where $j=2,3,4,5$. Further, for any $j\in \{1,2,3,4,5\}$ and $r\in D_j$, we have that $c(x)c(x^r)$ is a codeword of $\bm{C_1}$. Hence, from Theorem 3, we have $d^2_{1,j}\geq d_1=n_2$, i.e., $d_{1,j}\geq \lceil \sqrt{n_2} \rceil$, where $j=0,1,2,3,4,5$. By similar argument, we get $d_{2,j}\geq \lceil \sqrt{n_1} \rceil$, where $j=0,1,2,3,4,5$.$\square$

\noindent\textbf{Example 9} Let $q=2$ and $n_1=13,n_2=19$. Then the cyclic code over $\mathrm{GF}(q)$ with the generator polynomial $g(x)=\displaystyle\frac{x^n-1}{(x^{n_1}-1)d_0(x)}$ has parameters [247, 49, 19]. In this case, $d\geq \lceil \sqrt{n_2} \rceil=\lceil \sqrt{19} \rceil=5$, while the actual minimal distance is 19.

\begin{theorem}\label{thm6}
Suppose that $q \in D_0$. Let $\bm{C_{n_1,n_2}}$ and $d_{n_1,n_2}$ be defined as in Theorem \ref{thm4}. The cyclic code over $\mathrm{GF}(q)$ generated by the polynomial $g_{n_1,n_2,j}(x)=\displaystyle\frac{(x^n-1)(x-1)}{(x^{n_1}-1)(x^{n_2}-1)d_j(x)}$. The $\bm{C_{n_1,n_2,j}}$ has parameters $[n, n_1+n_2-1+$ \\ $\displaystyle\frac{(n_1-1)(n_2-1)}{6}$, $d_{n_1,n_2,j}]$, where $d_{n_1,n_2,j}\geq \lceil \sqrt{\min (n_1,n_2)} \rceil$, and $0\leq j \leq5$.
\end{theorem}
\noindent\textbf{Proof.} Let $j=0$ and $c(x)\in \mathrm{GF}(q)[x]/(x^n-1)$ be a codeword of Hamming weight $w$ in $\bm{C_{n_1,n_2,0}}$. Take $r\in D_1$, then $r^{-1}(\bmod n)\in D_5$ and $c(x^r)$ is a codeword of Hamming weight $w$ in $\bm{C_{n_1,n_2,5}}$, which implies that $d_{n_1,n_2,0}=d_{n_1,n_2,5}$. Take $r\in D_2$, then $r^{-1}(\bmod n)\in D_4$ and $c(x^r)$ is a codeword of Hamming weight $w$ in $\bm{C_{n_1,n_2,4}}$, which implies that $d_{n_1,n_2,0}=d_{n_1,n_2,4}$. By taking $r\in D_3$, $r\in D_4$ and $r\in D_5$ respectively, we can get $d_{n_1,n_2,0}=d_{n_1,n_2,3}$, $d_{n_1,n_2,0}=d_{n_1,n_2,2}$ and $d_{n_1,n_2,0}=d_{n_1,n_2,1}$ respectively. Further, for any $j\in \{1,2,3,4,5\}$ and $r\in D_j$, we have that $c(x)c(x^r)$ is a codeword of $\bm{C_{n_1,n_2}}$. Hence, from Theorem 4, we have $d_{n_1,n_2,j}^2\geq d_{n_1,n_2}=\min(n_1,n_2)$, i.e., $d_{n_1,n_2,j}\geq \lceil \sqrt{\min (n_1,n_2)} \rceil$, where $0\leq j \leq5$.$\square$

\noindent\textbf{Example 10} Let $q=2$ and $n_1=13,n_2=19$. Then the cyclic code over $\mathrm{GF}(q)$ with the generator polynomial $g(x)=\displaystyle\frac{(x^n-1)(x-1)}{(x^{n_1}-1)(x^{n_2}-1)d_1(x)}$ has parameters [247, 67, 13]. In this case, $d\geq \lceil \sqrt{\min (n_1,n_2)} \rceil=\lceil \sqrt{13} \rceil$, and the lower bound of $d$ is 4, while the actual minimal distance is 13.
\begin{theorem}\label{thm7}
Suppose that $q \in D_0$. Let $\bm{C_i}$ and $d_i$ be defined as in Theorem \ref{thm3}. Let $\bm{C_{i,j,h,t}}$ denote the cyclic code over $\mathrm{GF}(q)$ generated by the polynomial $g_{i,j,h,t}(x)=\displaystyle\frac{x^n-1}{(x^{n_i}-1)d_j(x)d_h(x)d_t(x)}$. The $\bm{C_{i,j,h,t}}$ has parameters $[n$, $n_i+\displaystyle\frac{(n_1-1)(n_2-1)}{2}$, $d_{i,j,h,t}]$, where $d_{i,j,h,t}\geq \lceil \sqrt{n_{i-(-1)^i}} \rceil$, $i=1,2$ , where $(j,h,t)\in \{(0,1,2),(0,1,5),(0,4,5),(1,2,3),(2,3,4),(3,4,5)\}$.$\square$
\end{theorem}
\noindent\textbf{Proof.} Let $i=1$, $j=0$, $h=1$, $t=2$ and $c(x)\in \mathrm{GF}(q)[x]/(x^n-1)$ be a codeword of Hamming weight $w$ in $\bm{C_{1,0,1,2}}$. Take $r\in D_1$, then $r^{-1}(\bmod n)\in D_5$ and $c(x^r)$ is a codeword of Hamming weight $w$ in $\bm{C_{1,0,1,5}}$, which implies that $d_{1,0,1,2}=d_{1,0,1,5}$. Take $r\in D_2$, then $r^{-1}(\bmod n)\in D_4$ and $c(x^r)$ is a codeword of Hamming weight $w$ in $\bm{C_{1,0,4,5}}$, which implies that $d_{1,0,1,2}=d_{1,0,4,5}$. By taking $r\in D_3$, $r\in D_4$ and $r\in D_5$ respectively, we can get $d_{1,0,1,2}=d_{1,3,4,5}$, $d_{1,0,1,2}=d_{1,2,3,4}$ and $d_{1,0,1,2}=d_{1,1,2,3}$ respectively. Further, for $r\in D_3$, we have that $c(x)c(x^r)$ is a codeword of $\bm{C_1}$. Hence, from Theorem 4, we have $d_{1,0,1,2}^2\geq d_1=n_2$. Hence, we have $d_{1,j,h,t}\geq \lceil\sqrt{n_2}\rceil$, where $(j,h,t)\in \{(0,1,2),(0,1,5),(0,4,5),(1,2,3),(2,3,4),(3,4,5)\}$. Similarly, we get $d_{2,j}\geq \lceil \sqrt{n_1} \rceil$, where
$(j,h,t)\in \{(0,1,2),(0,1,5),(0,4,5),(1,2,3),(2,3,4),(3,4,5)\}$.$\square$

\noindent\textbf{Example 11} Let $q=2$ and $n_1=13,n_2=19$. Then the cyclic code over $\mathrm{GF}(q)$ with the generator polynomial $g(x)=\displaystyle\frac{x^n-1}{(x^{n_1}-1)d_0(x)d_1(x)d_2(x)}$ has parameters [247, 49, 19]. In this case, $d\geq \lceil \sqrt{n_2} \rceil=\lceil \sqrt{19} \rceil=5$, while the actual minimal distance is 19.

\begin{theorem}\label{thm8}
Supposing $q \in D_0$. Let $\bm{C_{n_1,n_2}}$ and $d_{n_1,n_2}$ be defined as in Theorem \ref{thm4}. The cyclic code over $\mathrm{GF}(q)$ generated by the polynomial $g_{n_1,n_2,i,j,h}(x)=\displaystyle\frac{(x^n-1)(x-1)}{(x^{n_1}-1)(x^{n_2}-1)d_i(x)d_j(x)d_h(x)}$ has parameters $[n, n_1+n_2-1+\\\displaystyle\frac{(n_1-1)(n_2-1)}{2}, d_{n_1,n_2,i,j,h}]$, where $d_{i,j,h}\geq \lceil \sqrt{\min(n_1,n_2)}\rceil$, and $(i,j,h)\in \{(0,1,2),(0,1,5),(0,4,5),(1,2,3),(2,3,4),(3,4,5)\}$.
\end{theorem}
\noindent\textbf{Proof.} Let $i=0$, $j=1$, $h=2$ and $c(x)\in \mathrm{GF}(q)[x]/(x^n-1)$ be a codeword of Hamming weight $w$ in $\bm{C_{n_1,n_2,0,1,2}}$. Take $r\in D_1$, then $r^{-1}(\bmod n)\in D_5$ and $c(x^r)$ is a codeword of Hamming weight $w$ in $\bm{C_{n_1,n_2,0,1,5}}$, which implies that $d_{n_1,n_2,0,1,2}=d_{n_1,n_2,0,1,5}$. Take $r\in D_2$, then $r^{-1}(\bmod n)\in D_4$ and $c(x^r)$ is a codeword of Hamming weight $w$ in $\bm{C_{n_1,n_2,0,4,5}}$, which implies that $d_{n_1,n_2,0,1,2}=d_{n_1,n_2,0,4,5}$. By taking $r\in D_3$, $r\in D_4$ and $r\in D_5$ respectively, we can get $d_{n_1,n_2,0,1,2}=d_{n_1,n_2,3,4,5}$, $d_{n_1,n_2,0,1,2}=d_{n_1,n_2,2,3,4}$ and $d_{n_1,n_2,0,1,2}=d_{n_1,n_2,1,2,3}$ respectively. Further, for $r\in D_3$, we have that $c(x)c(x^r)$ is a codeword of $\bm{C_{n_1,n_2}}$. Hence, from Theorem 4, we have $d_{n_1,n_2,0,1,2}^2\geq d_{n_1,n_2}=\min(n_1,n_2)$, i.e., $d_{n_1,n_2,0,1,2}\geq \lceil\sqrt{\min(n_1,n_2)}\rceil$. Hence, we have $d_{n_1,n_2,i,j,h}\geq \lceil\sqrt{\min(n_1,n_2)}\rceil$, where $(i,j,h)\in \{(0,1,2),(0,1,5),(0,4,5),(1,2,3),(2,3,4),(3,4,5)\}$.$\square$


\section{Conclusion}
The generator polynomials of cyclic codes defined by two-prime generalized cyclotomic sequences of order 6 over $\mathrm{GF}(q)$ have been calculated in this paper. For the case that $q$ belongs to different generalized cyclotomic classes, we discussed in detail and gave the generator polynomial respectively. In the last section, we calculated the minimum distance of some cyclic codes and gave lower bounds of the minimum distance for some other cyclic codes. We will consider the application of some other cyclotomic sequences and generalized cyclotomic sequences of different periods and in different finite fields to form cyclic codes.

\end{document}